\documentclass[12pt,letterpaper,superscriptaddress]{JHEP3}
\pdfoutput=1
\usepackage{amssymb,amsmath,amsfonts,amstext,graphics}
\usepackage{cite}
\usepackage{graphicx}
\usepackage{epstopdf}
\input epsf.tex
\newdimen\tableauside\tableauside=1.0ex
\newdimen\tableaurule\tableaurule=0.4pt
\newdimen\tableaustep
\def\phantomhrule#1{\hbox{\vbox to0pt{\hrule height\tableaurule width#1\vss}}}
\def\phantomvrule#1{\vbox{\hbox to0pt{\vrule width\tableaurule height#1\hss}}}
\def\sqr{\vbox{%
  \phantomhrule\tableaustep
  \hbox{\phantomvrule\tableaustep\kern\tableaustep\phantomvrule\tableaustep}%
  \hbox{\vbox{\phantomhrule\tableauside}\kern-\tableaurule}}}
\def\squares#1{\hbox{\count0=#1\noindent\loop\sqr
  \advance\count0 by-1 \ifnum\count0>0\repeat}}
\def\tableau#1{\vcenter{\offinterlineskip
  \tableaustep=\tableauside\advance\tableaustep by-\tableaurule
  \kern\normallineskip\hbox
    {\kern\normallineskip\vbox
      {\gettableau#1 0 }%
     \kern\normallineskip\kern\tableaurule}%
  \kern\normallineskip\kern\tableaurule}}
\def\gettableau#1 {\ifnum#1=0\let\next=\null\else
  \squares{#1}\let\next=\gettableau\fi\next}
\newcommand{\gsim}{\lower.7ex\hbox{$\;\stackrel{\textstyle>}{\sim}\;$}}
\newcommand{\lsim}{\lower.7ex\hbox{$\;\stackrel{\textstyle<}{\sim}\;$}}
\def\OO{{\cal O}}
\def\LL{{\cal L}}
\def\HH{{\cal H}}

\def\eg{{\it e.g. }}
\def\ie{{\it i.e. }}

\newcommand{\TeV}{\,\text{ TeV}}
\newcommand{\GeV}{\,\text{ GeV}}
\newcommand{\MeV}{\,\text{ MeV}}
\newcommand{\keV}{\,\text{ keV}}

\newcommand{\half}{{\frac{1}{2}  }}

\newcommand{\hc}{\text{ h.c. }}
\newcommand{\identity}{{\rlap{1} \hskip 1.6pt \hbox{1}}}
\newcommand{\slD}{\!D\!\!\!\!\!\!\not\;\;}

\newcommand{\Tr}{{\text{ Tr }}}

\newcommand{\eff}{{\text{eff}}}

\newcommand{\be}{\begin{eqnarray}}
\newcommand{\ee}{\end{eqnarray}}
\newcommand{\bea}{\begin{eqnarray}}
\newcommand{\eea}{\end{eqnarray}}

\newcommand{\bef}{\begin{figure}[htbp]\begin{center}}
\newcommand{\eef}{\end{center}\end{figure}}

\newcommand{\pid}{{\pi_{\text{d}}}}
\newcommand{\pidn}[1]{{\pi_{\text{d}}^{(#1)}}}
\newcommand{\rhod}{{\rho_{\text{d}}}}
\newcommand{\td}{{\text{d}}} 
\newcommand{\alth}{\alpha_{\text{t}}}
\title{
\begin{flushright}
\mbox{\normalsize SLAC-PUB-14024}
\end{flushright}
\vskip 15 pt
The Cosmology of Composite Inelastic Dark Matter}
\author{Daniele S. M. Alves$^{a,b}$, Siavosh R. Behbahani$^{a,b}$, Philip Schuster$^{a}$, and Jay G. Wacker$^{a}$\\
$^a$ 
Theory Group,\\ 
SLAC National Accelerator Laboratory,\\
Menlo Park, CA 94025\\\\
$^b$Stanford Institute for Theoretical Physics,\\ 
Stanford University,\\ 
Stanford, CA 94306
}
\abstract{
Composite dark matter is a natural setting for implementing inelastic dark matter --- the $\OO(100\keV)$ mass splitting arises from spin-spin interactions of constituent fermions. In models where the  constituents are charged under an axial $U(1)$ gauge symmetry that also couples to the Standard Model quarks, dark matter scatters inelastically off Standard Model nuclei and can explain the DAMA/LIBRA annual modulation signal. This article describes the early Universe cosmology of a minimal implementation of a composite inelastic dark matter model where the dark matter is a meson composed of a light and a heavy quark. 
The synthesis of the constituent quarks into dark hadrons results in several qualitatively different configurations of the resulting dark matter composition depending on the relative mass scales in the system.  
}
\begin{document}

%%%%%%%%%%%%%%%%%%%%%%%%%%%%%%%%%%%%%%%%%%
\section{Introduction}

DAMA/LIBRA has an on-going  8.9$\sigma$ annual modulation signal of their single hit rate \cite{DAMA/NaI, DAMA/LIBRA,Bernabei:2010mq}.  If this signal is caused by dark matter scattering off their NaI crystals, then DAMA, when combined with other null direct detection searches, suggests that dark matter is scattering inelastically to an excited state split in energy by $\OO(100\keV)$\cite{TuckerSmith:2001hy , Weiner:2008, iDMConstraints1, iDMConstraints2}. Inelastic dark matter (iDM) is a testable scenario with XENON10 and ZEPLIN3 performing dedicated reanalyses of their data \cite{Angle:2009xb, Akimov:2010vk}.  IDM will be decisively confirmed or refuted in the XENON100, LUX, or CRESST science run this year \cite{Aprile:2009zzc,Fiorucci:2009ak}.

Composite inelastic Dark Matter (CiDM) is a model of iDM, where the $\mathcal{O}(100\keV)$ scale is dynamically generated by hyperfine interactions of a composite particle \cite{Alves:2009nf}. %Generically, the dark matter is a composite state of two constituents, one heavy and one light, bound by a confining gauge interaction. 
 (See \cite{Technibaryon1, Technibaryon2, Technibaryon3, Technibaryon4, CompositeDAMA1, CompositeDAMA2, CompositeDAMA3,CompositeDAMA4} for other examples of composite dark matter.)  
In the original CiDM model, the dark matter consists of a spin-0 meson, $\pid$, that has a single heavy constituent quark.  Adjacent in mass to $\pid$ is the spin-1 dark meson, $\rhod$,  and with the mass scales chosen in \cite{Alves:2009nf}, the mass splitting between $\pid$ and $\rhod$ is  $\OO(100\keV)$. The dark-matter origin in CiDM is non-thermal and requires  a primordial asymmetry between the number densities of heavy quarks to heavy antiquarks. 
This article addresses the early Universe cosmology of this minimal CiDM model, calculating the abundance of $\pid$ bound states relative to $\rhod$ states, dark baryon states or other exotic configurations of the dark quarks, as well as the direct detection properties of the various components.

The results of this article show that a wide range of final abundances of dark hadrons is possible, depending on the spectroscopy of the dark sector states.   Some spectra have dark matter dominantly in the form of $\pid$, while other spectra have dark baryons  dominating  the abundance.  In the latter case, where the dark baryons are the predominant dark matter constituent, the residual $\pid$ component interacts sufficiently strongly to account for DAMA's signal. In all cases, exotic dark matter components arise in CiDM with novel elastic scattering properties -- nuclear recoil events are suppressed at low-energy by a dark matter form factor.  The relative abundance of $\rhod$ to $\pid$ is typically $n_\rhod/n_\pid\sim 10^{-4}$.  Existing searches for down-scattering ($\rhod \rightarrow \pid$) in direct detection experiments are not constraining, but may be feasible in the near future. 

The organization of this article is as follows. Sec.~\ref{intro:summary} briefly reviews a specific implementation of CiDM presented in \cite{Alves:2009nf} that will be used throughout this analysis. 
The spectroscopy of this model is discussed in Sec.~\ref{Sec: Spectroscopy} and the synthesis of dark meson and dark baryon states in the early Universe in Sec.~\ref{Sec: DMSynth}.   Sec.~\ref{Sec: Results} is the primary result of this article and the qualitatively different results of the synthesis calculation are classified here. Sec.~\ref{Sec:Upscattered} addresses the upscattered fraction of dark pions and its implications for direct detection.  Sec.~\ref{Sec:Baryons} summarizes constraints on the dark baryon fraction that arise from direct dark matter searches such as CDMS and comments on the novel properties of the elastic scattering processes. 
Sec.~\ref{Sec:Conclusion} concludes with the outlook and further possibilities.

%%%%%%%%%%%
\subsection{Review of Axial CiDM Model}\label{intro:summary}

A wide variety of hidden sectors weakly coupled to the Standard Model have been considered in the literature, and the possibility that
dark matter is charged under hidden sector gauge forces has received considerable recent attention \cite{Nelson:2008hj,ArkaniHamed:2008qn,Pospelov:2008jd, HiddenValley, Cui:2009xq, Batell:2009vb, Finkbeiner:2009, Chang:2008, Ross:2009, Kaplan:2009ag, Morrissey:2009ur, Kaplan:2009de}.
In CiDM, dark matter is a bound state with  constituents charged under a hidden sector gauge group of the form $SU(N_c)\times U(1)_\td$, 
where the $U(1)_\td$ gauge boson kinetically mixes with Standard Model hypercharge gauge boson, and the $SU(N_c)$ group condenses at the GeV scale. 
Theories in which dark matter is charged under a new GeV-scale gauge group, as in CiDM, predict a variety of multi-lepton signals in $B$-factories and $\phi$-factories \cite{Essig:2009nc, Batell:2009yf, :2009pw, :2009qd, Bossi:2009uw}, low-energy upcoming fixed-target experiments \cite{Bjorken:2009mm, Batell:2009di, Essig:2010xa, Reece:2009un}, and distinctive astrophysical signatures \cite{Schuster:2009fc, Schuster:2009au, Meade:2009mu}. 

The Lagrangian of the axial CiDM model in \cite{Alves:2009nf} is given by
\begin{equation}
\LL = \LL_{\text{SM}} + \LL_{\Psi}+\LL_{\text{Gauge}}+ \LL_{\text{Higgs}}
\end{equation}
where
\begin{eqnarray}
\label{eq:mix}
\nonumber
&\LL_{\text{Gauge}}=- \frac{1}{2} \Tr G_{\td}^2  - \frac{1}{4} F_{A_\td}^2 + \half\epsilon F_{A_\td}^{\mu\nu}B_{\mu\nu} \\
\nonumber
&\LL_{\Psi} = \bar{L} i \slD L + \bar{L}^c i \slD L^c +\bar{H} i \slD H + \bar{H}^c i \slD H^c\\
 &\LL_{\text{Higgs}} = |D_\mu \phi|^2 - \lambda( |\phi|^2 - v_\phi^2)^2
 + (y_L \phi L L^c + y_H \phi^* H H^c+\hc\!),
\end{eqnarray}
 and the gauge charges are
\begin{eqnarray}
\label{table charges}
\begin{array}{|c||c|c|c|}
\hline
& SU(N_c)& U(1)_A&U(1)_{H-L}\\
\hline\hline
L& \tableau{1}& 1&-1\\
L^c &\bar{\tableau{1}}& 1&+1\\
H& \tableau{1} & -1&+1\\
H^c & \bar{\tableau{1}} &-1&-1\\
\hline
\phi& \identity& -2&0\\
\hline
\end{array}
\end{eqnarray}
The fermion sector has  an $U(2)_{\text{left}} \times U(2)_{\text{right}}$ chiral flavor symmetry, broken down to a vector-like $U(1)_H\times U(1)_L$ by Yukawa interactions with a dark Higgs boson, $\phi$.   The vacuum expectation value of $\phi$, $\langle \phi \rangle = v_\phi$, causes the Abelian gauge field  to acquire a mass $m_{A_\td} = 2\sqrt{2} g_\td v_\phi$ and the fermions to pair into a light and a heavy Dirac fermion, 
\begin{eqnarray}
\Psi_L = (L, \bar{L}^c) \qquad \Psi_H = (H, \bar{H}^c)
\end{eqnarray}
with masses $m_L = y_L v_\phi$ and  $m_H=y_H v_\phi$, respectively.   In this minimal model, $m_H$ is near the electroweak scale while the other quark has a mass $m_L$ at or beneath the confinement  scale, $\Lambda_\td\sim \OO(100\MeV - 10\GeV)$.    

In analogy to the Standard Model without weak interactions, the $U(1)_H\times U(1)_L$ flavor symmetry renders the lightest mesons and baryons charged under this symmetry stable.  
The lightest stable bound state is a meson containing a single $\Psi_H$ and a $\bar{\Psi}_L$, which will be denoted as $\pid$. This dark meson is the dark matter candidate in \cite{Alves:2009nf}. 
Because the constituents are fermions, $\pid$ is paired with a vector state, $\rhod$. The spin-spin interactions of the constituents generate a hyperfine splitting
\begin{eqnarray}
\label{Eq: HF}
\Delta E_{\text{hyperfine}}= m_{\rhod} - m_{\pid} \simeq \left\{
  \begin{aligned}
\frac{\Lambda_\td^2}{ m_H} , \quad m_L< \Lambda_\td \qquad    \\
\frac{\alth^4m_L^2}{ m_H} , \quad m_L> \Lambda_\td \qquad
  \end{aligned}
  \right.
\end{eqnarray}
where $\alth$ is the $SU(N_c)$ 't Hooft coupling, and the parameters are chosen such that the splitting is at the scale $\mathcal{O}(100\keV)$  suggested by DAMA/LIBRA.  The hyperfine structure is described in more detail in Sec. \ref{FSSpec}.

Another key feature of this model is that the dark gauge boson, $A_\td^\mu$, couples axially to the dark quarks.   
The coupling between the $A^\mu_\td$ and the dark mesons are constrained by parity and all leading order scattering channels are forbidden but the $\pid -\rhod$ transition \cite{Alves:2009nf}. 
$A_\td^\mu$ mixes with the Standard Model photon and mediates dark meson/baryon scattering off SM nuclei.  In particular, $\pid$ up-scattering can explain the DAMA/LIBRA annual modulation signal. CiDM direct detection phenomenology is discussed in detail in \cite{Alves:2009nf,  Lisanti:2009vy, Lisanti:2009am}.

The mass of the dark Higgs, $\phi$, is radiatively unstable and introduces a second gauge hierarchy problem to the dark matter - Standard Model theory.   The solution to the Standard Model's gauge hierarchy problem may also solve this new hierarchy problem.  
Supersymmetric extensions of these models may solve both hierarchy problems at once, and may introduce new phenomena into the theory. 
For instance, if the only communication of supersymmetry breaking to the dark sector occurs through kinetic mixing, then the dark sector may be nearly supersymmetric, resulting in nearly supersymmetric bound states \cite{Rube:2009yc,Herzog:2009fw}.
These susy bound states may have different scattering channels and the phenomenology may
be different than minimal CiDM \cite{Behbahani:2010}.

%%%%%%%%%%%%%%%%%%%%%%%%%%%%%%%%%%%%%%%%%%%%%%%
\section{Dark Matter Synthesis Prelude: Hadron Spectroscopy}\label{Sec: Spectroscopy}

The origin of the dark matter abundance in CiDM is non-thermal, possibly a baryogenesis-like process that generates a non-zero $H-L$ number,  see \cite{Baryogenesis1,Baryogenesis2,Baryogenesis3,Baryogenesis4,Baryogenesis5,Baryogenesis6,Baryogenesis7,Baryogenesis8,Baryogenesis9,Baryogenesis10,Baryogenesis11,Baryogenesis12} for examples.  Most often, particularly at large $N_c$, no net dark baryon number is generated, resulting in a zero $H+L$ number.  This article addresses how the residual $H$ and $L$ asymmetry is configured at late times, \ie whether the states are in dark mesons or in dark baryons.

At temperatures above $m_H$, there is a thermal bath of gluons and free quarks and anti-quarks. When the temperature drops below $m_H$, the heavy dark quarks rapidly annihilate, and only the asymmetric abundance of $H$ remains. The non-relativistic heavy quarks are much heavier than the confining scale and can form Coulombic bound states through antisymmetric color channels. 
Light $\bar{L}$ antiquarks have binding energies comparable to the confinement scale and therefore never bind at temperatures $T \gsim\Lambda_\td$.
At confinement, the heavy quarks are dominantly screened by light antiquarks to form $SU(N_c)$ singlets. 

The evolution of the $H$ and $L$ asymmetry after confinement is dominantly determined by the spectroscopy of  the bound state systems.  After the heavy quark mass, the largest energy scale in the problem is the Coulombic binding energy between the heavy quarks.
The energetics of these heavy quark configurations guide the synthesis of baryon states through hadron binding reactions.
The analysis of Sec. \ref{Sec: DMSynth} illustrates the qualitative behavior of the synthesis of the $H$ and $L$ asymmetry in the early Universe. 
  The fraction of dark mesons synthesizing into dark baryons depends sensitively on the masses of the light hadrons and the results are parameterized by the mass of the lightest hadron.  
  
%%%%%%%%%%
\subsection{Coulombic $\Psi_H$ Binding Energies}
\label{Sec: Coulomb} 

The Coulombic binding energies of collections of heavy quarks is relevant for
dark matter synthesis.  Collections of  $n_H\leq N_c$ heavy quarks can  form quasi-Coulombic bound states that are deeply bound relative to the confinement scale when $ m_H \gg \Lambda_\td$.  The deepest bound states of $n_H$ heavy quarks are antisymmetric color configurations.    This section computes the binding energy of multiple $H$ bound states
using a variational approach.

The Coulomb potential between two heavy quarks, $i$ and $j$, in an antisymmetric
color configuration is given by \cite{DeRujula:1975ge,Banks:1975zw,Raby:1979my,Georgi:1990um}
\begin{eqnarray}
\label{Eq: Coulomb}
V_{ij}(r_{ij}) = \frac{g^2}{4\pi}\frac{1}{2r_{ij}}
C(2,1,1)
=  -\frac{\alth}{2r_{ij}}\frac{(1 + N_c^{-1})}{N_c},
\end{eqnarray}
where   $g$ is the ordinary gauge coupling and the running 't Hooft coupling is defined as
\begin{eqnarray}
\alth(\mu)=\frac{N_c g^2(\mu)}{4\pi}
\end{eqnarray}
and $C(n_1,n_2,n_3)$ is the following combination of Casimirs for
$n_i$-rank antisymmetric tensors of $SU(N_c)$
\begin{eqnarray}
\label{Eq: Casimirs}
C(n_1,n_2,n_3) = C_2(n_1)-C_2(n_2)-C_2(n_3).
\end{eqnarray}
The Casimirs for rank $n$ antisymmetric tensors are
\begin{eqnarray}
C_2(n) =C_2(\bar{n})=\frac{n(N_c-n)}{2}\left(1 + \frac{1}{N_c}\right)
\end{eqnarray}
where $\bar{n} \equiv N_c -n$ in the above expression.

Notice that the potential in Eq. \ref{Eq: Coulomb} for combining fundamentals into antisymmetric tensors is attractive but  $N_c$ suppressed for fixed 't Hooft coupling in contrast to a quark-antiquark potential.

The confinement scale, $\Lambda_\td$, is determined by the 't Hooft coupling evaluated at a scale $\mu=m_H$
\begin{eqnarray}
\Lambda_\td=m_H \exp\left(-\frac{6\pi}{(11-2/N_c)\alth(m_H)}\right),
\end{eqnarray}
for the minimal model with a single light flavor. 

The total Hamiltonian for a system of $n_H$ heavy quarks is
\begin{eqnarray}
\HH = \sum_{i=1}^{n_H} \frac{p_i^2}{2m_H} + \frac{1}{2}\sum_{i\ne j} V_{ij}(r_{ij}).
\end{eqnarray}
In the ground state configuration, the color indices of the quarks are antisymmetric with all other quantum numbers being symmetrized. In particular, all the quarks will have the same spatial wave function in the ground state. 
The ansatz for the bound state system will be
\begin{eqnarray}
\Psi^{(n_H)}(x_1,...,x_{n_H})= \prod_{i=1}^{n_H} \psi(x_i),
\end{eqnarray}
where $\psi(x)$ is a hydrogen-like ground state wave function of the form
\begin{eqnarray}
\psi(x) = \left(\frac{\mu_*^{3}}{8\pi}\right)^\half e^{-\mu_* r/2},
\end{eqnarray}
where $\mu_*$ is the variational parameter.
Evaluating the expectation value of $\HH$ gives an estimate of the binding energy 
\begin{eqnarray}
\langle \HH (\mu_*)\rangle  = n_H \left( \frac{\mu_*{}^2}{8m_H} - \frac{(1+N_c^{-1})}{4N_c}(n_H-1)\alth(\mu_*) \mu_* \right).
\end{eqnarray}
Minimizing $\langle\HH \rangle$ with respect to $\mu_*$ gives the inverse Bohr radius
for the $n_H$ heavy quark system:
\begin{eqnarray}
\mu_*(n_H)\simeq (n_H -1) \frac{\alth(\mu_*)}{N_c} m_H (1+ N_c^{-1}),
\end{eqnarray}
and binding energies of
\begin{eqnarray}
E_{B}(n_H) = 
   n_H(n_H-1)^2 E_{B0}\quad, \qquad E_{B0} = \left( \frac{  \alth(\mu_*)}{2 N_c}\right)^2 \frac{ m_H}{8}(1 + N_c^{-1})^2 
\label{eq:binding}
\end{eqnarray}
where the running of the 't Hooft coupling has been neglected in the differentiation.
For the case of $n_H=2$ the binding energy reduces to
\begin{eqnarray}
\label{Eq: first binding}
E_{B}(2)=  \half  \left( \frac{ \alth(\mu_*)}{2N_c}\right)^2  \frac{ m_H}{2},
\end{eqnarray}
precisely the analogue of the Rydberg constant for the potential in (\ref{Eq: Coulomb}).

The estimates of (\ref{eq:binding}) are crucial for two dominant rearrangement processes that occur
\begin{eqnarray}
\label{rearrangement reactions}
\Psi^{(n)}+\Psi^{(n')} \rightarrow \Psi^{(n+n')} +X_\td\quad,
\qquad
\Psi^{(n)}+\Psi^{(n')} \rightarrow \Psi^{(n+n'-m)} +\Psi^{(m)} .
\end{eqnarray}
Ignoring the effects of running on the 't Hooft coupling, the energy released in these reactions is
\begin{eqnarray}
\label{Eq: Energy Released}
\nonumber
Q^{2\rightarrow 1}_{n,n'} &=& E_B(n+n') - E_B(n) - E_B(n')-m_{X_\td}\\
Q^{2\rightarrow 2}_{m;n,n'} &=& E_B(n+n'-m)+ E_B(m)- E_B(n) - E_B(n').
\end{eqnarray}
The energy release from the first few reactions are tabulated in Table \ref{Tab: Energy Release}.
The first reaction releases the least energy and is a potential
bottleneck in the chain reaction that takes $N_c \Psi^{(1)} \rightarrow \Psi^{(N_c)}$.

\begin{table}
\begin{center}
\begin{tabular}{|c||c|c||c|}
\hline
$n+n'$& $n$& $n'$& $Q^{2\rightarrow 1}_{n,n'}$\\
\hline
2& 1&1& $2\,E_{B0}-m_{X_\td}$\\
3& 2& 1& $10\, E_{B0}-m_{X_\td}$\\
4 & 2&2 & $32\,E_{B0}-m_{X_\td}$\\
4& 3 &1& $24\,E_{B0}-m_{X_\td}$\\
\hline
\end{tabular}\;
%%\\
%%\vspace{0.2in}
\begin{tabular}{|c|c||c|c||c|}
\hline
$n+n'-m$&m& $n$& $n'$& $Q^{2\rightarrow 2}_{ m;n,n'}$\\
\hline
3 & 1&2&2 & $8\,E_{B0}$\\
4&1& 3 &2& $22\,E_{B0}$\\
\hline
\end{tabular}
\caption{
\label{Tab: Energy Release}
The energy release in the first $2\rightarrow 1$ and $2\rightarrow 2$ processes.
}
\end{center}
\end{table}

Eqs. \ref{eq:binding} and \ref{Eq: Energy Released} are the main results of this section. They will completely determine whether heavy quark synthesis into bound states takes place before confinement. The role of light states (in particular light quarks) during the unconfined phase is irrelevant due to the large gluon entropy at such temperatures. After confinement, when the $\Psi^{(n_H)}$ states are color-neutralized by light quarks, the estimates of (\ref{eq:binding}) and (\ref{Eq: Energy Released}) are still expected to hold, since they physically correspond to deeply bound states of heavy quarks and are typically an order of magnitude larger than $\Lambda_\td$ and the binding energies of light quarks. The energetics of the rearrangement reactions (\ref{rearrangement reactions}) after confinement will be completely determined by the results of (\ref{eq:binding}) and the parametric dependence on the mass of the lightest state in the spectrum, that is typically present among the final states $X_\td$ (more about that on Sec.~\ref{Sec: PostSynth}).

\subsection{Hadronization of Bound States}
\label{Sec: Hadronization}

There are two ways of constructing color singlets from $n_H$ heavy quarks in an antisymmetric color configuration: mesons and baryons.   For mesons, the color is neutralized by $n_H$ light antiquarks in an antisymmetric color configuration, while for baryons the color is neutralized by $(N_c -n_H)$ light quarks in a color antisymmetric state.
In QCD, the analogues to these multi-heavy quark systems would be di-heavy baryon and tetraquark states of the form
\begin{eqnarray}
ccq, cbq, bbq\qquad   c c\bar{q} \bar{q}', cb \bar{q} \bar{q}', b b \bar{q} \bar{q}'
\end{eqnarray}
where $q$ and $q'$ are light-flavored quarks \cite{DeRujula:1976qd,Jaffe:1976ig,Jaffe:1976ih,Gelman:2002wf,DelFabbro:2004ta,Janc:2004qn}. Only the $ccq$ has been observed, but there is an ongoing program to discover the rest of these states.   

In the dark sector of CiDM, the corresponding states will be denoted by $\pidn{n_H}$ for meson states\footnote{
$\pidn{n}$ refers to all spin configurations of $n$ $\Psi_H$ and $n$ $\bar{\Psi}_L$ dark quarks. Specifically, $\pidn{1}$ refers to both $\pid$ and $\rhod$.}
 with $n_H$ heavy quarks and by $B^{(n_H)}$ for baryon states with $n_H$ heavy quarks. 
For convenience, $B^{(N_c)} \equiv B_H$ and $B^{(0)}\equiv B_L$. 

Due to the difference in the number of light quarks necessary to hadronize the color of $\pidn{n_H}$ and $B^{(n_H)}$, these states will have different masses.
Using a constituent picture of the light quarks, the masses of mesons and baryons are given by
\begin{eqnarray}
\label{Eq: Full Bind}
\nonumber
m_{\pidn{n_H}} &=& n_H m_H - E_{B}(n_H) + m_\eff (n_H) + m_{\text{spin}}(n_H)\\
m_{B^{(n_H)}} &=&  n_H m_H - E_B(n_H)  + m_\eff(N_c-n_H) + m_{\text{spin}} (N_c-n_H).
\end{eqnarray}

In  Eq. \ref{Eq: Full Bind},  $m_\eff(n_L)$ parametrizes the constituent mass of the light quarks confined in a bag of size $ \Lambda_\td$
%%
%where  $m_\eff(n_L)$ corresponds to the constituent mass of the light quarks in the hadron and $m_{\text{spin}}(n_L)$ is the spin-spin interaction of the light quarks. The former is parametrized by e%ffective mass of $n_L$ light quarks confined in a bag of size $ \Lambda_\td$,
%%
\begin{eqnarray}
\label{Eq: ConstituentMass}
m_\eff(n_L) = n_L(1 + \hat{m}_0(n_L)) \Lambda_\td.  
\end{eqnarray}
In principle this leading order effective mass could be $n_L$-dependent, due to changes in  the light quark wavefunctions.    For simplicity, this article will take $\hat{m}_0(n_L)=0$.

 The term $m_{\text{spin}}(n_H)$ in Eq. \ref{Eq: Full Bind}  arises from the spin-spin interaction amongst the light quarks and a constituent model of this interaction gives
 \begin{eqnarray}
 m_{\text{spin}}(n_L) \propto  C(2,1,1)  \sum_{i\ne j} \langle \vec{S}_i \cdot \vec{S}_j\rangle .
 \end{eqnarray}
Evaluating the expectation values above for a totally spin-symmetric state, one finds
 \begin{eqnarray}
m_{\text{spin}}(n_L) = \hat{m}_1(n_L)  n_L   \frac{n_L-1}{4N_c} (1 + N_c^{-1}) \Lambda_\td.
 \end{eqnarray}
 Using the $\Lambda_b - \Sigma_b$ system in QCD one infers: $\hat{m}^{\Sigma_b}_1(2) = 1.14$.
 This article will use $\hat{m}_1(n_L) =1.0$.

Both the constituent masses and spin-spin interactions cause the mesons to be lighter than the baryons when $n_H < \half N_c$ and the baryons to be lighter than the mesons in the complimentary case.

The most relevant reactions in the early Universe that synthesize stable $B^{(n)}$ and $\pidn{n}$ states have rates that are exponentially sensitive to total binding energy differences.  The estimates of the binding energies have an uncertainty of $\OO(\Lambda_\td)$ that has been absorbed into the unknown constants in Eq.~\ref{Eq: Full Bind}.   
In practice, the binding energy differences are larger than $\Lambda_\td$ by roughly an order of magnitude so that the Coulombic spectroscopy dominantly determines the synthesis of $\pidn{1}$ into other species of dark hadrons.

\subsection{Hyperfine Structure Spectroscopy}
\label{FSSpec}

The stable dark hadrons have a large degeneracy of their ground states from the suppressed spin-spin interactions of the heavy quarks with the light quarks.
Both in dark mesons and dark baryons, same-flavor quarks have antisymmetrized colors and symmetrized spins.
Therefore, heavy quarks are in a $S_H= n_H/2$ spin configuration while light quarks are in a $S_L=n_H/2$ spin configuration for mesons and $S_L= (N_c-n_H)/2$ spin configuration for baryons.
The resulting range of spins for dark mesons and baryons is, respectively,
\begin{eqnarray}
0\le j_{\pid^{(n_H)}}\le n_H\quad,\qquad
\Big|\half N_c -n_H\Big|\le  j_{B^{(n_H)}}\le \half N_c  .
\end{eqnarray}
Hence, the degeneracy of a bound state containing $n_H$ heavy quarks (either mesonic or baryonic) is equal to $n_H+1$. Spin-spin interactions break this degeneracy and introduce hyperfine splittings, resulting in the lowest spin configuration being the lowest energy configuration. Such splittings can be very suppressed with respect to the dark matter mass and offer a natural mechanism to generate the $\OO(100\keV)$ scale suggested by DAMA.

The spin-spin coupling of heavy and light degrees of freedom splits the ground state degeneracy of hadrons.   This is studied for Standard Model $b$ hadrons in
\cite{Jenkins:1996de,Jenkins:2007dm,Lewis:2008fu}.  The quark chromomagnetic moment is suppressed by its constituent mass, $m_*$, and is given by
\begin{eqnarray}
|\vec{\mu}| =  \frac{ g(\mu)}{ m_*} \simeq \begin{cases} 
\frac{g(\mu)}{m} & m \gg \Lambda_\td\\
 \sqrt{\frac{4\pi}{N_c}} \frac{1}{\kappa \Lambda_\td} & m\ll \Lambda_\td
\end{cases}
\end{eqnarray}
where the coupling is evaluated at the effective Bohr radius for the entire, color-neutral bound state. For $m\ll \Lambda_\td$, $m_* \rightarrow \Lambda_\td$, $\alth \rightarrow 1$, and $\kappa$ is introduced to fix the relationship.
The first order correction to the energy levels due to spin-spin coupling of heavy and light degrees of freedom is 
\begin{eqnarray}
\label{Eq: Spin-Spin 0}
E_{\text{Hyperfine}}\simeq  -C(0, n_H, \bar{n}_H) |\vec{\mu}_H| |\vec{\mu}_L|\langle\vec{S}_H\cdot\vec{S}_L\rangle|\psi(0)|^2,
\end{eqnarray}
where $\vec{S}_H$ and $\vec{S}_L$ are the spin operators for the collection of heavy and light quarks, respectively.  
The energy splittings can be evaluated using
\begin{eqnarray}
\vec{S}_H\cdot\vec{S}_L = \half(S^2 - S_H^2 - S_L^2),
\end{eqnarray}
where $\vec{S} = \vec{S}_H + \vec{S}_L$.  The $S_H^2$ and $S_L^2$ terms do not induce mass splittings.
The color factor $C(0,n_H, \bar{n}_H)$ in Eq. \ref{Eq: Spin-Spin 0} is defined in Eq. \ref{Eq: Casimirs} and is given by
 \begin{eqnarray}
C(0,n_H,\bar{n}_H)=  n_H(N_c-n_H)\left( 1 +\frac{1}{N_c}\right).
 \end{eqnarray}
 Note that the color expression above applies to both mesons and baryons.
 
 Finally, $\psi(0)$ in Eq. \ref{Eq: Spin-Spin 0} is the ground state wave function (\textit{i.e.}, $n=l=0$) for the light quarks evaluated at the origin and is roughly
 \begin{eqnarray}
 |\psi(0)|^2\simeq \frac{1}{4\pi} \begin{cases}
 1/a_B^3= (\alth m_L)^3 & m_L \gg \Lambda_\td  \\
(\kappa \Lambda_\td)^3 & m_L \ll \Lambda_\td.
 \end{cases}
\end{eqnarray}
For states with multiple light quarks, \ie all states but $\pidn{1}$ and $B^{(N_c-1)}$, 
there could be significant multi-body effects that could change the wave functions.
In particular, QCD baryons have a hyperfine splitting smaller than that of mesons by a factor of 
roughly three.  For bottom hadrons
\begin{eqnarray}
\frac{m_{B^*} - m_{B}}{S^2_{B^*}- S^2_B} 
\frac{S^2_{\Sigma_b^*} - S^2_{\Sigma_b}}{ m_{\Sigma_b^*}- m_{\Sigma_b}} = \left(\frac{\kappa^b_{n_L=1}}{\kappa^b_{n_L=2}}\right)^2
\simeq 3.23 .
\end{eqnarray}
Remarkably similar ratios hold for charm hadrons and even strange and light flavored hadrons.
This is evidence for $n_L$ dependence in $\kappa$ and can be interpreted as $\hat{m}_0$ from Eq. \ref{Eq: ConstituentMass} being
\begin{eqnarray}
 \hat{m_0}(n_L) =  \frac{1}{\kappa_{n_L}}-1.
\end{eqnarray}
 This dependence in $\kappa_{n_L}$ should be universal and applicable to both baryons and mesons.

Combining the results for $m_L\ll \Lambda_\td$ gives
\begin{eqnarray}
E_{\text{Hyperfine}}\simeq \kappa_{n_L}^2\; \langle\vec{S}_H\cdot\vec{S}_L\rangle \; n_H\left(1-\frac{n_H}{N_c}\right) \frac{ \Lambda_\td^2}{m_H}\left(1  + \OO( N_c^{-1})\right)
\end{eqnarray}
where $n_L=n_H$ for $\pidn{n_H}$ states and $n_L=N_c-n_H$ for $B^{(n_H)}$ states.
The energy splitting between the ground state and the first excited state for dark mesons is
\begin{eqnarray}
\Delta E_{\pidn{n_H}}\simeq
\kappa_{n_H}^2\; n_H\left(1-\frac{n_H}{N_c}\right) \frac{\Lambda_\td^2}{m_H} , \quad m_L\ll \Lambda_\td .
\end{eqnarray}
For baryons the hyperfine splitting of the ground state is
\begin{eqnarray}
\label{hf}
\Delta E_{B^{(n_H)}}\simeq \kappa_{N_c-n_H}^2\; n_H\left(1-\frac{n_H}{N_c}\right) \left(\Big|\half N_c -n_H\Big|+1\right)\frac{\Lambda_\td^2}{m_H}, \quad m_L\ll \Lambda_\td.
\end{eqnarray}
Fitting expression (\ref{hf}) to the hyperfine splitting of $B$ and $B^*$ in the Standard Model, one finds 
\begin{eqnarray}
\kappa^b_1 \simeq 1.35 
\end{eqnarray}
for $\Lambda_{\text{QCD}} =350 \MeV$ and $m_b= 4.5 \GeV$.  Throughout this paper this constant will be taken to be $\kappa_1=1.0$. 

QCD indicates $\kappa_n < \kappa_1$, potentially making the smallest hyperfine splitting be in the $\pidn{n}$ meson, rather the $\pidn{1}$ meson.  This opens up an alternate explanation of DAMA:  that it is the multi-heavy quark mesons that are responsible for scattering. 

\subsection{Light States}
\label{Sec: Light States}

Just like hybrid mesonic and baryonic bound states, unstable states are in thermal equilibrium with the thermal bath at early times.
The unstable spectrum consists of states that carry no net $H$ and $L$ quantum numbers. There are no Goldstone bosons, since this is a single flavor theory at confinement and the global $U(1)_\text{F}$ is anomalous. Therefore the mass of the lightest state is of the order of the confinement scale and could either be a light meson or a glueball state \cite{Donoghue:1980hw,Jaffe:1985qp, Strassler:2009ji}.  This section describes the unstable, $\OO(\Lambda_\td)$ mass states,  estimating their lifetimes and briefly describing their cosmology.

\begin{table}
\begin{center}
\begin{tabular}{|c|c|c|}
\hline
Name & $J^{PC}$& Decay Mode\\
\hline
$\eta_\td$ & $0^{-+}$ & $4\ell$\\
$\sigma_\td$ & $0^{++}$ & 4$ \ell$\\
$\omega_\td$ & $1^{--}$ & 2$\ell$\\
\hline
\end{tabular}
\caption{\label{Tab: Light Mesons} Examples of light mesons}
\end{center}
\end{table}

Long lived states are of potential concern to Standard Model BBN. We will show here, however, that a general depletion mechanism causes the abundance of quasi-stable states to be negligible by the time BBN begins. Among the long lived states is the  $\eta_\td \sim \bar{\Psi}_L\gamma_5 \Psi_L$  meson, which decays through the operator
\begin{eqnarray}
\LL_{\eta_\td} = \frac{\alpha_\td}{4\pi}\frac{N_c}{f_{\pid}}\eta_\td F_{\td\,\mu\nu}\tilde{F}_\td^{\mu\nu}
\end{eqnarray}
to four Standard Model leptons through two off-shell $A^\mu_\td$'s. 
The resulting decay rate scales as \cite{Essig:2009nc}:
\begin{eqnarray}
\Gamma(\eta_\td \rightarrow4l)\sim\frac{\alpha_\td^2}{64\pi^3f_{\pi_\td}^2}m_{\eta_\td}^3\left(\frac{1}{4\pi}\alpha\epsilon^2\frac{m_{\eta_\td}^4}{m^4_{A_\td}}\right)^2,
\end{eqnarray}
giving $\eta_\td$ a typical lifetime of 
\begin{eqnarray}
\tau_{\eta_\td}=10^{8}~\text{s} \left( \frac{\GeV}{m_{\eta_\td}} \right)^9 \left( \frac{m_{A_\td}}{\GeV} \right)^8 \left( \frac{4\times10^{-12}}{\epsilon^2\alpha_\td} \right)^2.
\end{eqnarray}
Scalar glueballs with $J^{PC}=0^{-+}$ mix with $\eta_\td$ mesons and inherit the same decay channels. $J^{PC}=0^{++}$ scalar glueballs, on the other hand, mix with $\sigma_\td$ and can decay to either on-shell or off-shell $A_\td$'s or light dark mesons. Their lifetime is expected to be no longer than the lifetime of the $0^{-+}$ states.

Among the short-lived states are $\omega_\td$ vector mesons and $J^{PC}=1^{--}$ glueballs that can decay through mixing with $\omega_\td$. Their estimated lifetime is \cite{Essig:2009nc}:
\begin{eqnarray}
\tau_{\omega_\td}&\sim& \left(  \frac{4\pi}{3}\epsilon^2\alpha\alpha_D \right)^{-1}\frac{(m^2_{\omega_\td}+m^2_{A_\td})^2}{m^5_{\omega_\td}}
\end{eqnarray}
\begin{eqnarray}
\sim5\times10^{-12}\text{ s}\left(\frac{\GeV}{ m_{\omega_\td}}\right)^5\left(1+\frac{m^2_{\omega_\td}}{m^2_{A_\td}}\right)^2\left( \frac{m_{A_\td}}{\GeV} \right)^4 \left( \frac{4\times10^{-12}}{\epsilon^2\alpha_\td} \right).
\end{eqnarray}

The existence of short lived states in the spectrum is a leaky bottom mechanism for quasi-stable flavorless hadrons. These long lived states are depleted by scattering into shorter lived ones, {\it e.g.}, $\eta_\td+ X \rightarrow \omega_\td +X'$, with cross sections set by $\Lambda_\td$:
\begin{eqnarray}
\langle\sigma v\rangle_{\eta_\td+ X \rightarrow \omega_\td +X'}\simeq 1/\Lambda_\td^2 .
\end{eqnarray}
The residual abundance of quasi-stable particles by the time Standard Model BBN begins is
\begin{eqnarray}
\label{longlived}
\zeta_{\eta_\td}\equiv \frac{n_{\eta_\td}}{s}m_{\eta_\td} \sim 10^{-18}\GeV,
\end{eqnarray}
for $m_{\eta_\td}\sim\Lambda_\td$. 

The most stringent constraints for a late decaying relic $X$ with hadronic branching fraction $B_{\text{had}}=1$ are set by $^3\text{He/D}$ \cite{Kawasaki:2004qu}:
\begin{eqnarray}
\label{BBNbound}
\zeta_X\equiv \frac{n_X}{s}m_X\lsim10^{-14}\GeV.
\end{eqnarray}
The relic abundance (\ref{longlived}) is several orders of magnitude below the upper limit (\ref{BBNbound}). This is a conservative bound because the decay products of the dark hadrons are predominantly leptons or photons, which alter the primordial abundances of elements less than hadronic decays. Therefore,  light states have a negligible effect on BBN.

%%%%%%%%%%%%%%%%%%%%%%%%%%%%%%%%%%%
\section{Dark Matter Synthesis and Evolution}
\label{Sec: DMSynth}

The evolution of the $H$ and $L$ asymmetry in the early Universe is divided into three stages ordered by the temperature relative to the confinement scale: above, at, or below. 
Starting at temperatures close to $m_H$, the thermal abundance of dark $H$-quarks is exponentially suppressed and only the non-thermal component is left to synthesize into composite states. 
At temperatures above the confinement scale, perturbative techniques are applicable to estimate the cross sections for rearrangement reactions
because  $m_H\gg\Lambda_\td$.
For temperatures above confinement, the light quarks can be ignored throughout, and included during confinement to screen the color charge of the heavy quark bound states. Below confinement the light quarks can be treated as spectators. Assuming that approach is justified, the computation is reasonably accurate up to $\OO(1)$-factors.

%%%%%%%%%%%%%%%%%
\subsection{Preconfinement Synthesis}
\label{Pre Synth}

At high temperatures, $T \gsim \Lambda_\td$, the non-Abelian gauge dynamics is unconfined and quasi-perturbative.   Light quarks are always relativistic and do not form bound states. Below temperatures $T\sim \OO(m_H)$,  heavy quarks can form Coulombic bound states through antisymmetric color channels (as described in Sec. \ref{Sec: Coulomb}) by reactions of the kind
\begin{eqnarray}
\label{Eq: reaction}
\Psi^{(n)} + \Psi^{(n')}\rightarrow \Psi^{(n+n')} + g\quad,&\qquad&Q=Q^{2\rightarrow 1}_{n,n'} \quad (m_{X_\td}=0) \\
\Psi^{(n)} + \Psi^{(n')}\rightarrow \Psi^{(n+n'-m)} + \Psi^{(m)}\quad,  &\qquad& Q = Q^{2\rightarrow 2}_{m;n,n'}
\end{eqnarray}
where $\Psi^{(n)}$ is an antisymmetric bound state of $n$ heavy quarks ($2 \le n\le N_c$), $g$ is a dark gluon  and the energy releases, $Q^{2\rightarrow 1}_{n,n'}$ and $Q^{2\rightarrow 2}_{m;n,n'}$ are given in Eq.~\ref{Eq: Energy Released} and tabulated in Table~\ref{Tab: Energy Release} with $m_{X_\td}=m_g=0$.  

The capture cross section for reactions (\ref{Eq: reaction}) is given by \cite{Kolb:1990vq},
\begin{eqnarray}
\label{Eq: cross-section}
\langle \sigma v\rangle _{nn'}=\frac{4\pi^2}{\mu^2_{nn^{\prime}}}\frac{\alth}{N_c}  \frac{Q^{2\rightarrow 1}_{n,n'}}{ (3\mu_{nn^{\prime}}T)^{1/2}},
\end{eqnarray}
where $\mu_{nn^{\prime}}$ is the reduced mass of $\Psi^{(n)} + \Psi^{(n^{\prime})}$, $(3\mu_{nn^{\prime}}T)^{1/2}$ accounts for Sommerfeld enhancement, and  $\alth(\mu_*)$ is the 't Hooft coupling evaluated at the Bohr radius of the bound state $\mu_* =  \alth m_H/2N_c$.

Each reaction of the type (\ref{Eq: reaction}) contributes to the Boltzmann equation for the number density of $\Psi^{(n)}$, $\mathsf{n}_n$, as\footnote{with appropriate factors of $2$ included whenever $n=n'$.}
\begin{eqnarray*}
\frac{d\mathsf{n}_n}{dt} + 3H\mathsf{n}_n = \ldots -\langle\sigma v\rangle _{nn^{\prime}}
\times \left[\mathsf{n}_n \mathsf{n}_{n'} - \mathsf{n}_{n+n'}\left(\frac{T}{2\pi}\frac{m_n m_{n'}}{m_{n+n'}} \right)^{3/2}\exp(-Q^{2\rightarrow 1}_{n,n'}/T) \right] + \ldots
\end{eqnarray*}
The exponential factor in the Boltzmann equation above accounts for the large gluon entropy that prevents heavy quark bound state formation down to temperatures
\begin{eqnarray}
T_*\simeq\frac{E_{B}(2)}{|\ln(Y_H)|},
\end{eqnarray}
where $Y_H \equiv \mathsf{n}_H/s$. Plugging in Eq.~\ref{Eq: first binding}, we see that strong gluon entropy dissociation is effective down to temperatures when confinement takes place for:
\begin{eqnarray}
\frac{m_H}{\GeV}\lsim\frac{16N_c^2}{\alpha_t(\mu_*)^2}|\ln(Y_H)|\Lambda_d\qquad\qquad
\end{eqnarray}
\begin{eqnarray}
\qquad\qquad\simeq\frac{16N_c^2}{\alpha_t(\mu_*)^2}\left[\ln\left(\frac{m_H}{\GeV}\right)+21\right]\Lambda_d.
\end{eqnarray}

For $\Lambda_d \gsim \OO(100 \MeV)$ and $N_c=4$ that corresponds to $m_H \lsim \OO(10\TeV)$. Thus for the parameter space favored by DAMA/LIBRA, there is no pre-confinement bound state formation.

%%%%%%%%%%%%%%%%%
\subsection{Synthesis at Confinement}
\label{Sec: Synth Con}

It was shown in Sec.~\ref{Pre Synth} that dissociation due to relativistic dark gluons inhibits bound state formation down to temperatures $T\simeq\Lambda_\td$.  At confinement, this dissociation shuts off since the theory acquires a mass gap.   The light quarks and gluons form massive dark hadrons and proceed to decay promptly (Sec. \ref{Sec: Light States}).  

The formation of heavy quark bound states is suppressed since those are dilute at the time of confinement and there is a strong nucleation of light quark-antiquark pairs that screen the color charge of free heavy quarks.  So, confinement will preferentially lead to formation of $\pid$ dark mesons over higher-$n_H$ dark hadrons.

It is possible, however, to estimate the abundance of $n_H=2$ dark hadrons formed at confinement by simple combinatorics.  If two heavy quarks are apart by a distance smaller that $\Lambda_\td^{-1}$ at confinement, the light-quark screening effect will not operate and the two heavy quarks will bind.  Computing the probability that that will happen gives an estimate for the ratio of $n_H=2$ dark hadrons over $\pid$ dark mesons produced at confinement
\begin{eqnarray}
\frac{n_{\Psi^{(2)}}  }{n_{\pid}} \sim \frac{n_{\pid}}{\Lambda_\td^3} \sim 10^{-10}\left(\frac{100\GeV}{m_H}\right).
\end{eqnarray}
These bound states of two heavy quarks will either be screened by two light anti-quarks to form a $\pidn{2}$ dark meson, or by $(N_c-2)$ light quarks to form a $B^{(2)}$ dark baryon. The later is Boltzmann-suppressed over the former for $N_c > 4$ since it has $(N_c-4)$ more light constituents.

The formation of dark baryons at confinement  depends sensitively on $ T_{\text{conf}}$. Using the leading term from Eq.~\ref{Eq: ConstituentMass}
\begin{eqnarray}
\frac{n_{B^{(n)}}}{n_{\pidn{n} }}
 \simeq \exp\left( -(N_c-2n) \frac{ \Lambda_\td}{T_{\text{conf}}}\right) .
\end{eqnarray}
Using QCD as a guide, $T_{\text{conf}} \simeq 0.3 \, \Lambda_\td$, leads to
\begin{eqnarray}
\frac{n_{B^{(1)}}}{n_{\pidn{1} }} = 3\times 10^{-4} 
\end{eqnarray}
for $N_c=4$. The $n_H=1$ dark baryons can annihilate with $B_{\bar{L}}$ into $\pidn{1}+ X_\td$.  The residual $B^{(1)}$ abundance that does not annihilate will quickly synthesize into $B_H$.

%%%%%%%%%%%%%%%%
\subsection{Postconfinement Synthesis}
\label{Sec: PostSynth}

After the transition from the unconfined to the confined phase, the dark quarks hadronize dominantly into $\pidn{1}$  dark mesons with a suppressed fraction in $B^{(1)}$ dark baryons.  
Heavier dark hadron synthesis occurs through
\begin{eqnarray}
\pidn{1}+\pidn{1}\rightarrow \pidn{2}+\pidn{0}\quad,\qquad B^{(1)}+ \pidn{1} \rightarrow B^{(2)}+ \pidn{0},
\end{eqnarray}
 where $\pidn{0}$ is the lightest dark hadron, such as a glueball or light meson, with mass $m_{\text{light}}\simeq \OO(\Lambda_\td)$. Once $\pidn{2}$ mesons start forming, a chain of reactions can occur that will ultimately lead to formation of dark baryons $B_H$.
 
As an illustration consider an $N_c=4$ theory. The reaction chain down to heavy dark baryon synthesis is:
\begin{subequations}
\label{all reactions}
\begin{align}
\label{Eq: first reaction}
\pidn{1}+\pidn{1} &\rightarrow \pidn{2}+\pidn{0}& Q =& 2 E_{B0} - m_{\text{light}}\\
\label{Eq: first baryon reaction}
B^{(1)}+\pidn{1} &\rightarrow B^{(2)} + \pidn{0} & Q = &2 E_{B0} + (m_{\text{eff}}(2)  - m_{\text{light}})\\
\label{Eq: second reaction}
 \pidn{1}+\pidn{2}&\rightarrow\pidn{3}+\pidn{0}& Q = &10 E_{B0} - m_{\text{light}}&\\
  \label{Eq: third reaction}
  \pidn{2}+\pidn{2}&\rightarrow\pidn{3} + \pidn{1}& Q=&8E_{B0} \\
  \label{Eq: no endo1}
 \pidn{1}+\pidn{3}&\rightarrow B_H+B_{\bar{L}}& Q= &24 E_{B0}\\
  \label{Eq: no endo2}
  \pidn{2}+\pidn{2}&\rightarrow B_H+B_{\bar{L}}& Q= &32 E_{B0}.
  \end{align}
 \end{subequations}
 As a combination of the mass gap in the theory and the fact that the heavy quarks in $\pidn{2}$ are not particularly deeply bound, the first reaction (\ref{Eq: first reaction}) becomes endothermic for a large fraction of the parameter space.  On the other hand, the $B^{(1)}$ reaction (\ref{Eq: first baryon reaction}) is always exothermic; therefore, dark baryons formed at confinement process more efficiently than dark mesons.   The results listed in Sec.~\ref{Sec: Results} assume that all hybrid dark baryons efficiently process into $B_H$.
 
 Reactions that have hybrid baryons in the final state
 \begin{eqnarray}
 \label{Eq: first reaction baryon}
\pidn{n}+ \pidn{n'} \rightarrow B^{(n+n')} + B_{\bar{L}}
 \end{eqnarray}
 are energetically disfavored compared to their mesonic counterparts,
 \begin{eqnarray}
 \label{Eq: first reaction baryon}
\pidn{n}+ \pidn{n'} \rightarrow \pidn{n+n'} + \pidn{0},
 \end{eqnarray}
 and have negligible contribution to the dark matter synthesis.
 
% Reactions that  form hybrid baryons, such as 
 %%
 %\begin{eqnarray}
 %\label{Eq: first reaction baryon}
%\pidn{1}+ \pidn{1} \rightarrow B^{(2)} + B_{\bar{L}},
 %\end{eqnarray}
% %
%are kinematically disfavored compared with (\ref{Eq: first reaction}) because they contain more constituent light quarks in their final products: three constituent light quarks Eq. \ref{Eq: first reaction} %versus five in Eq. \ref{Eq: first reaction baryon}.  Using Eq. \ref{Eq: Full Bind},
%the difference in energy released between the reactions is
%%
%\begin{eqnarray}
%\nonumber
%Q_{ \pidn{1}\pidn{1} \rightarrow  \pidn{2}\pidn{0}} - Q_{\pid\pid\rightarrow B^{(2)}  \bar{B}_L}&=&
%\Big(2N_c -4 - \hat{m}_{\text{light}}\\
%&&\hspace{0.25in} + \hat{m}_0(N_c-2) +\hat{m}_0(N_c)-  \hat{m}_0(2) \Big)\Lambda_\td .
%\end{eqnarray}
%%
%The baryonic reaction is suppressed at temperatures beneath $\Lambda_\td$.
%The next-to-last step of synthesis,  when forming $\pidn{N_c-1}$ versus $B^{(N_c-1)}$, is a borderline situation when $\hat{m}_{\text{light}} \simeq 2$.  $\hat{m}_{\text{light}} \simeq 2$ is suggested %by the constituent picture of light quarks and the baryonic reaction may not be particularly suppressed relative to the mesonic reaction.  However, if  $\hat{m}_{\text{light}} \simeq 2$, as Fig. \ref{Fig: %Spectrum} shows, there is very little synthesis and the exclusion of the next-to-last baryonic reaction is irrelevant in practice. 

Heavy quark binding in the rearrangement reactions (\ref{all reactions}) requires large momentum transfer $p_{\text{min}}=m_Hv\simeq\sqrt{2m_HT}\gg\Lambda_\td$, and hence it is expected that the cross sections are controlled by perturbative heavy quark dynamics, as it was during pre-confinement (\ref{Eq: cross-section}). However, there are two important differences between pre- and post-confinement processing:
\begin{enumerate}
\item There is no Sommerfeld enhancement because the lightest degrees of freedom are heavier than the typical momentum transfer in such reactions: $m_{\text{light}}> p_{\text{min}}=m_Hv$. This amounts to a reduction of the post-confinement cross section by a factor of $(\alth/v)$ relative to the pre-confinement cross section.
\item If $m_{\text{light}}$ is heavier than the $\pidn{2}$ binding energy, the first reaction (\ref{Eq: first reaction}) is \textit{endothermic}, which further suppresses post-confinement synthesis by an additional Boltzmann factor $e^{-(m_{\text{light}}-E_B(2))/T}$ in the thermally averaged cross sections.
\end{enumerate}
Therefore, post-confinement cross sections for heavy hadron processing compares to pre-confinement heavy quark binding (\ref{Eq: cross-section}) as:
 \begin{eqnarray}
 \langle \sigma v\rangle _{\text{post}}= \left(\frac{\alth(p_{\text{min}})}{v}\right)^{-1} e^{-(m_{\text{light}}-\Delta E_B)/T}  \langle \sigma v\rangle _{\text{pre}}.
\end{eqnarray}

For the range of parameters where these reactions are endothermic, processing will be already frozen out by the time of confinement. This demonstrates that no heavy baryons are formed for $m_{\text{light}}\sim\Lambda_\td\gsim E_{B}(2)$, or:
\begin{eqnarray}
\label{Eq: no synth}
\Lambda_\td\gsim \left(\frac{\alth(\mu_*)}{4N_c}\right)^2m_H.
\end{eqnarray}
However, that is not the only possibility. As one explores other ranges for the confinement scale $\Lambda_\td$ and the mass of the lightest state $m_\text{light}$, the full numerical solution to the Boltzmann equations reveals that the synthesis of CiDM can allow for much richer range of compositions detailed in Sec. \ref{Sec: Results}. 

\subsection{Synthesis Results}
\label{Sec: Results}

\begin{figure}[htbp]
\begin{center}
\includegraphics[width=2.8in]{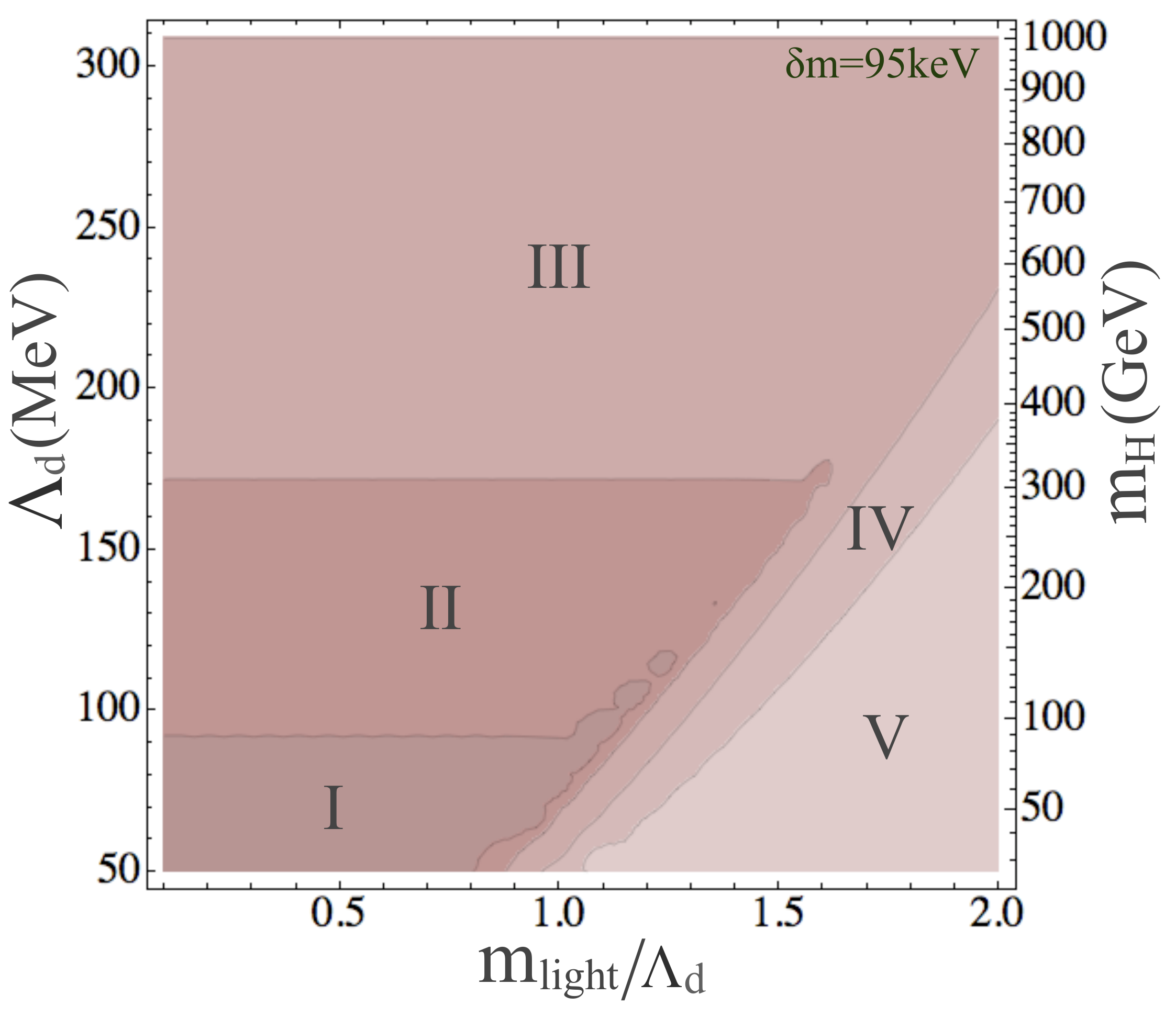}
\includegraphics[width=3.2in]{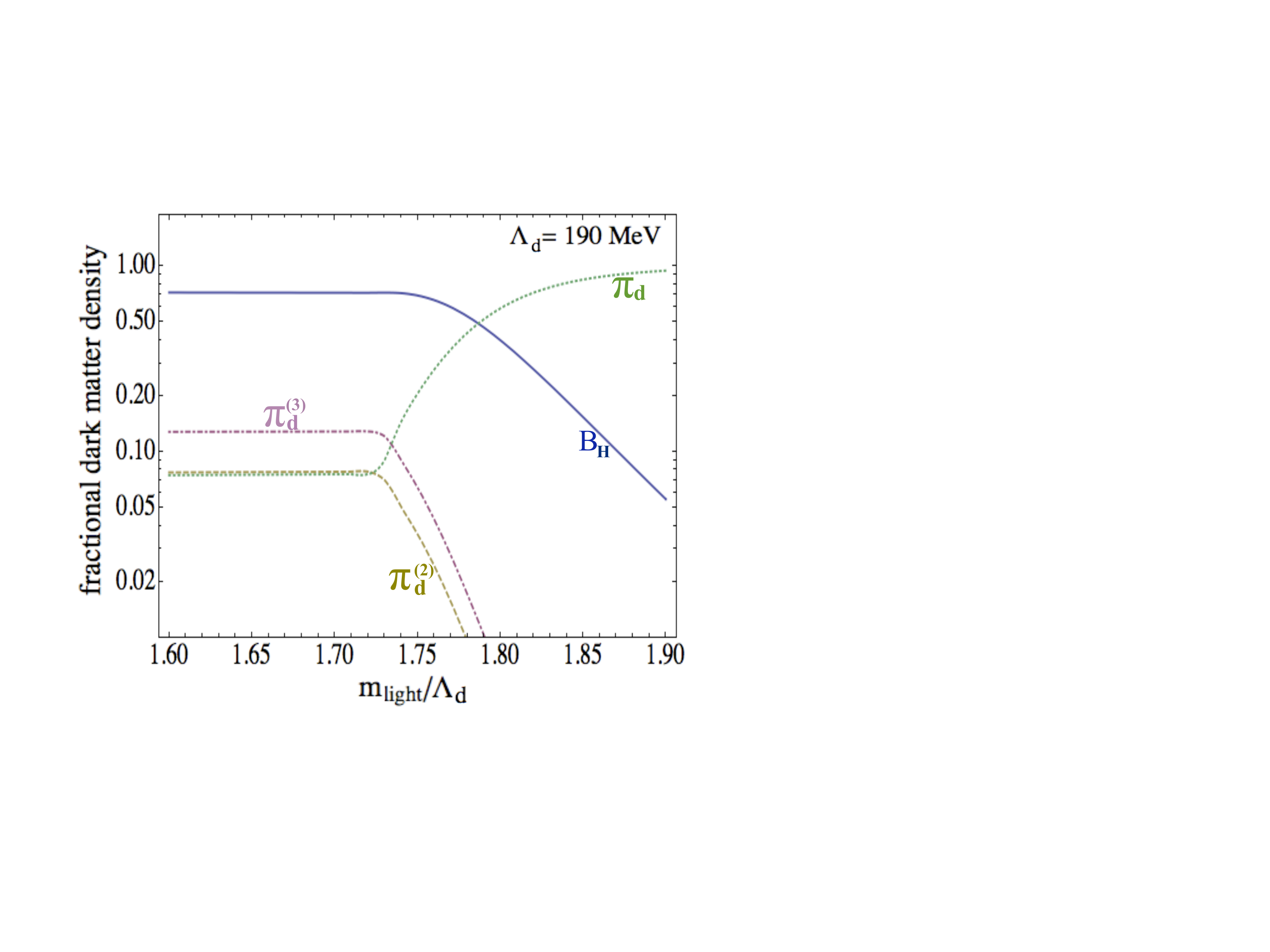}
\caption{\label{Fig: Spectrum} {\it Left:}~Dark matter synthesis as a function of the confining scale in the dark sector $\Lambda_\td$ and the mass of the lightest state $m_{\text{light}}$ for an $N_c=4$ dark sector. The inelastic splitting is $\delta m = 95\keV$. The different regions are described in the text and in Table 3. {\it Right:}~Synthesis for $\Lambda_\td$ fixed at $190\MeV$ and $\delta m = 95 \keV$.}
\end{center}
\end{figure}

Fig.~\ref{Fig: Spectrum} illustrates the CiDM synthesized spectrum as a function of the $\Lambda_\td-m_{\text{light}}$ parameter space for the case $N_c=4$.
A general classification into five regions of different qualitative behavior is possible:
\begin{description}
\item[Region I: Complete] Synthesis is efficient and the CiDM composition is dominated by heavy baryons $B_H$.  DAMA's signal may arise from inelastic scattering of the highly suppressed dark meson components, particularly of $\pidn{3}$ if $\kappa_{n_L}$ is a decreasing function of $n_L$.    The viability of this region requires that the heavy and light baryons are not visible in direct detection experiments. That will be further discussed in Sec.~\ref{Sec:Baryons}.
\item[Region II: Nearly Complete] Heavy baryons are the dominant component, with a few percent of the dark matter density in the form of dark pions $\pidn{1}$, $\pidn{2}$ and $\pidn{3}$.   This region is not as extreme as Region I for CiDM. As with Region I, DAMA's signal may arise from scattering of the sub-dominant dark meson component.  
\item[Region III: Incomplete] Synthesis results in a democratic abundance with comparable mass densities for all for states $\pidn{1}$, $\pidn{2}$, $\pidn{3}$ and $B_H$.  Here the dominant {\it number density} of dark matter is the $\pidn{1}$ state and there are other components of the dark matter to discover.
\item[Region IV: Arrested] The complete synthesized and unsynthesized components, $B_H$ and $\pid$, share comparable mass densities, with  a few percent in the form of exotic pions, $\pidn{2}$ and $\pidn{3}$.   The first step of the synthesis chain, $\pidn{1}\pidn{1} \rightarrow \pidn{2} \pidn{0}$ is the bottleneck much like deuterium formation slows BBN in the Standard Model. It only occurs for a brief period, but once the $\pidn{2}$ has formed, it processes quickly into $B_H$.
\item[Region V: Inhibited] The first step of the synthesis chain is strongly supressed and the CiDM composition is dominated by $\pidn{1}$.  Region V is the cosmology taken in \cite{Alves:2009nf}. The heavy baryon component mostly arises through the primordial $B^{(1)}$ formation described in Sec. \ref{Sec: Synth Con}.
\end{description}
A quantitative description of the abundances in each one of these regions is summarized in Table \ref{Tab: Regions}.

\begin{table}[ht]
\begin{center}
\begin{tabular}{|c||c|c|c|c|}
\hline
Region & $\rho_\pidn{1}/\rho_{\text{DM}}$\vspace{0.02in} &  $\rho_\pidn{2}/\rho_{\text{DM}}$ & $\rho_\pidn{3}/\rho_{\text{DM}}$ &$ \rho_{B_H}/\rho_{\text{DM}}$ \\
\hline\hline
I & $10^{-4}-0.1\%$ & $10^{-4} - 0.2\% $& $10^{-3} - 0.9\% $&$ >99\% $\\
\hline
II & $0.1\% - 4\% $&$ 0.2\% - 5\%$ &$0.9\% - 11\%$ & $80\% - 99\%$ \\
\hline
III & $4\% - 57\% $&$ 5\% - 24\% $& $11\% - 17\% $&$ 9\% - 80\%$ \\
\hline
IV & $57\% - 99\%$ & $<5\%$ & $<5\%$ & $1\% - 30\%$ \\
\hline
V & $>99\%$ &$ <10^{-5} $&$ <10^{-5}$ &$ <1\%$ \\
\hline
\end{tabular}
\caption{
\label{Tab: Regions}
The relations on the fractional mass densities that define the regions of dark matter synthesis in Fig.~1.
}
\end{center}
\end{table}

The region of interest for DAMA/LIBRA realizes the full variety of dark matter synthesis possibilities.
In all cases, $\pid$ is a good candidate for explaining the DAMA/LIBRA signal through inelastic kinematics, but signals from other dark matter components may be important. This will be discussed further in Sec. \ref{Sec:Baryons}. 
 
%%%%%%%%%%%%%% 
\section{The Excited Fraction of CiDM}
\label{Sec:Upscattered}

Sec.~\ref{Sec: DMSynth} demonstrated that for a large range of the parameter space favored by DAMA/LIBRA \cite{Alves:2009nf,Essig:2009nc,Lisanti:2009am} CiDM has a dark matter halo dominated by $n_H=1$ mesons, with subdominant components in other configurations.  This section describes how the spin 1, $\rhod$ dark mesons become depopulated.    The discussion of this section is couched in ``Region V'' taking in the minimal CiDM model of \cite{Alves:2009nf} with a $\pid$ dominated halo.  This analysis can be generalized to any of the synthesis scenarios considered in 
Sec. \ref{Sec: Results} and the results do not significantly change.

The nearly degenerate $n_H=1$ meson spin states $\pid$ and $\rhod$ are equally populated at high temperatures. When kinetic decoupling between these two spin states takes place, the excited dark matter fraction, $n_{\rhod}/n_{\pid}$, is frozen in. If the excited state $\rhod$ is cosmologically long lived, the fractional number density will be constrained from direct detection of $\rhod\rightarrow \pid$ de-excitation in nuclear scattering to be $n_{\rhod}/n_{\pid} \lsim 10^{-2}$\cite{Finkbeiner:2009, Batell:2009vb}.

These bounds are relevant whenever kinetic mixing of $U(1)_\td$ with hypercharge is the only channel for $\rhod$ to decay. Then, the only kinematically allowed decays are to $\pid$ plus photons or neutrinos. The decay to photons is highly suppressed (a loop-induced 3-photon decay). The kinetic mixing decay to neutrinos is suppressed by an additional factor of $\epsilon q^2/m_{A_\td}^2$ in the amplitude. The resulting $\rhod$ lifetime is much longer than the age of the Universe in these cases \cite{Finkbeiner:2009, Batell:2009vb}.

\subsection{Early Time Depopulation}

Down scattering constraints from long-lived excited states are a common challenge to all iDM models coupled to the Standard Model only through kinetic mixing of $U(1)_\td$ with hypercharge. Kinetic decoupling of the spin states usually occurs before $T\simeq 1 \MeV$, leading to an $\OO(1)$ fractional number density $n_{\rhod}/n_{\pid} $. Strongly coupled composite dark matter models exhibit an elegant way to depopulate the excited states. 

The reactions that keep the dark meson spin states in kinetic equilibrium have large cross sections set by the size of the composite
\begin{eqnarray}
\label{Eq: De-excitation Cross Section}
\langle\sigma v\rangle_{\rhod\rhod\rightarrow \pid\pid} \simeq \frac{\hat{r}^2_0}{\Lambda_{\text{d}}^2}.
\end{eqnarray}
where $\hat{r}_0$ is an $\OO(1)$ constant that parameterizes the quasi-elastic  scattering cross section.
These reactions will freeze out when
\begin{eqnarray}
\label{SpinTempFreezeOut}
n_{\rhod} \langle\sigma v\rangle_{\rhod\rhod\rightarrow \pid\pid}~\simeq~3 e^{-\delta m/T_*}n_{\pid} \hat{r}_0^2\Lambda_{\text{d}}^{-2}~\lsim~H(T_*),
\end{eqnarray}
where $\delta m\sim\OO(100\keV)$ is the energy splitting between $\rhod$ and $\pid$, $T_*$ is the freeze out temperature and $H(T_*)$ is the Hubble expansion rate at freeze-out. With $\Lambda_{\text{d}}^2\sim  m_H \delta m/\kappa_1^2$, Eq. \ref{SpinTempFreezeOut} implies that the primordial up-scattered fraction $n_{\rhod}/n_{\pid} \simeq 3 e^{-\delta m/T_*}$ is completely determined by the dark pion mass:
\begin{eqnarray}
\frac{ n_{\rhod}/n_{\pid}}{|\ln n_{\rhod}/n_{\pid}|}
\approx
5 \times 10^{-6}\; \kappa_1^2\hat{r}_0^2\; \left(\frac{m_{\pid}}{100 \GeV} \right)^2.
\end{eqnarray}

In other synthesis regions where the $\pidn{1}$ abundance is suppressed, \eg Region I or II,  the depopulation if $\rhod$ is more effective  because the dominant de-excitation interaction is
\begin{eqnarray}
B_H + \rhod \rightarrow B_H + \pid.
\end{eqnarray}
This interaction has roughly the same cross section as (\ref{Eq: De-excitation Cross Section}), but $B_H$ acts as a catalyst for de-excitation resulting in
\begin{eqnarray}
\label{Eq: Catalyzed SpinTempFreezeOut}
n_{B_H}\hat{r}_0^2\Lambda_{\text{d}}^{-2}~\lsim~H(T_*)
\quad \Rightarrow  \quad 
T_*\simeq  2~\text{eV}  
\left(\frac{m_\pid}{100 \GeV}\right)^2 \!\!
\left(\frac{\delta m}{100 \keV}\right)\!\! \left(\frac{ N_c/4}{\kappa_1^2 \hat{r}_0^2}\right)
\end{eqnarray}
%%
%%
%\begin{eqnarray}
%\label{Eq: Catalyzed SpinTempFreezeOut}
%\frac{\xi(B_H) \rho_{\text{DM}}  \hat{r}_0^2}{m_{B_H}\Lambda_\td^2 v}  \lsim H
%\quad \Rightarrow  \quad 
%T_*\simeq  1 \keV  
%\left(\frac{m_\pid}{100 \GeV}\right)^{\!\frac{5}{3}} \!\!
%\left(\frac{\delta m}{100 \keV}\right)^{\!\frac{2}{3}}\!\! \left(\frac{ N_c}{\xi(B_H)\hat{r}_0^2 \kappa_1^2}\right)^{\!\!\frac{2}{3}}
%\end{eqnarray}
%%
which severely depletes the $\rhod$ population.

\subsection{Late Time Up Scattering}

After structure formation begins, the velocity of the dark matter increases allowing for late-time up scattering of $\pid$ into $\rhod$ in dark matter halos. The excited state can then be repopulated to observable levels today. The up scattering reaction
\begin{eqnarray}
\pid+\pid\rightarrow\rhod+\rhod
\end{eqnarray}
is endothermic, so its cross section scales as
\begin{eqnarray}
\sigma_{\pid\pid\rightarrow\rhod\rhod}\sim\frac{1}{|q|^2} \sqrt{ 1 - \frac{ \delta m}{ E} } \simeq\frac{1}{m_{\pid} \delta m }.
\end{eqnarray}
The number of up scattering collisions since structure started forming, about $\tau_{\text{struct}}\sim10^{10}\, $yr ago, is roughly:
\begin{eqnarray}
\label{eq: late upscat}
n_{\pid}^{\text{halo}}\langle\sigma v\rangle\tau_{\text{struct}}\sim 10^{-1}v_{\text{virial}} \left(\frac{100\GeV}{m_{\pid}} \right)^2 \left(\frac{100\keV}{\delta m} \right)
\end{eqnarray}
with $v_{\text{virial}}\simeq 10^{-3}$.  For $m_{\pid}\simeq 100\GeV$, a fraction $\OO(10^{-4})$ of the $\pid$ have up-scattered during the entire age of the Universe. Up-scattering from reactions of $\pid$ off nuclei through $A_\td^\mu$-exchange can generate a similar $\rhod$ abundance.

Eq.~\ref{eq: late upscat} demonstrates that, even though up-scattering during structure formation does not surpasses the primordial excited fraction for $m_{\pid}\gsim\OO(100\GeV)$, it is the primary up-scattering effect for $m_{\pid}<\OO(100\GeV)$, reaching the current upper bound $n_{\rhod}/n_{\pid}\lsim\OO(10^{-2})$ for dark pion masses as low as $\OO(10\GeV)$.

\subsection{Down Scattering Discovery}

Although the predicted CiDM excited fraction $n_{\rhod}/n_{\pid}$ is safely below the current upper bound, it is in the reach of discovery of the next generation of direct detection experiments.  In order to quantify that, in what follows we compute the detection rate of the $\rhod$ component as it down-scatters off of target detector nuclei.

The down-scattering interaction is determined by the operator $(2g_\text{d}/\Lambda_{\text{d}})\pid\partial_\nu\rhod_\mu F^{\mu\nu}_{A_{\text{d}}}$, where $F^{\mu\nu}_{A_{\text{d}}}$ mixes with Standard Model hypercharge $B^{\mu\nu}$ (see Eq. \ref{eq:mix}). Its differential rate is $1/3$ of the differential rate for up-scattering \cite{Alves:2009nf},
\be
\frac{d\sigma}{dE_R}=\frac{1}{3}\left(\frac{m_N}{\Lambda_{\text{d}}}\right)^2 \frac{4\alpha Z^2}{f_\eff^4}\frac{1}{v_{\text{rel}}^2}E_R |F(E_R)|^2.
\ee
Here $E_R$ is the nucleus recoil energy, $m_N$ and $Z$ are its mass and atomic number, respectively, and $v_{\text{rel}}$ is the $\rhod$-nucleus relative velocity. The Helm nuclear form factor $|F(E_R)|^2$ accounts for loss of coherence scattering off of the entire nucleus at large recoil, and it is given by
\begin{equation}
|F(E_R)|^2=\left(\frac{3j_1(|q|r_0)}{|q|r_0}\right)^2e^{-s^2|q|^2},
\end{equation}
where $s=1\text{ fm}$, $r_0=\sqrt{r^2-5s^2}$, $r=1.2A^{1/3}\text{ fm}$, and $|q|=\sqrt{2m_NE_R}$ is the momentum transfer. Finally, $f_\eff$ is defined by:
\begin{equation}
\frac{1}{f_{\text{eff}}^4}\equiv\frac{\epsilon^2 g_\text{A}^2}{m_{A_\td}^4},
\end{equation}
where $m_{A_\td}$ is the mass of the $U(1)_\td$ gauge boson.

\begin{figure}[htbp]
\begin{center}
\includegraphics[width=4.5in]{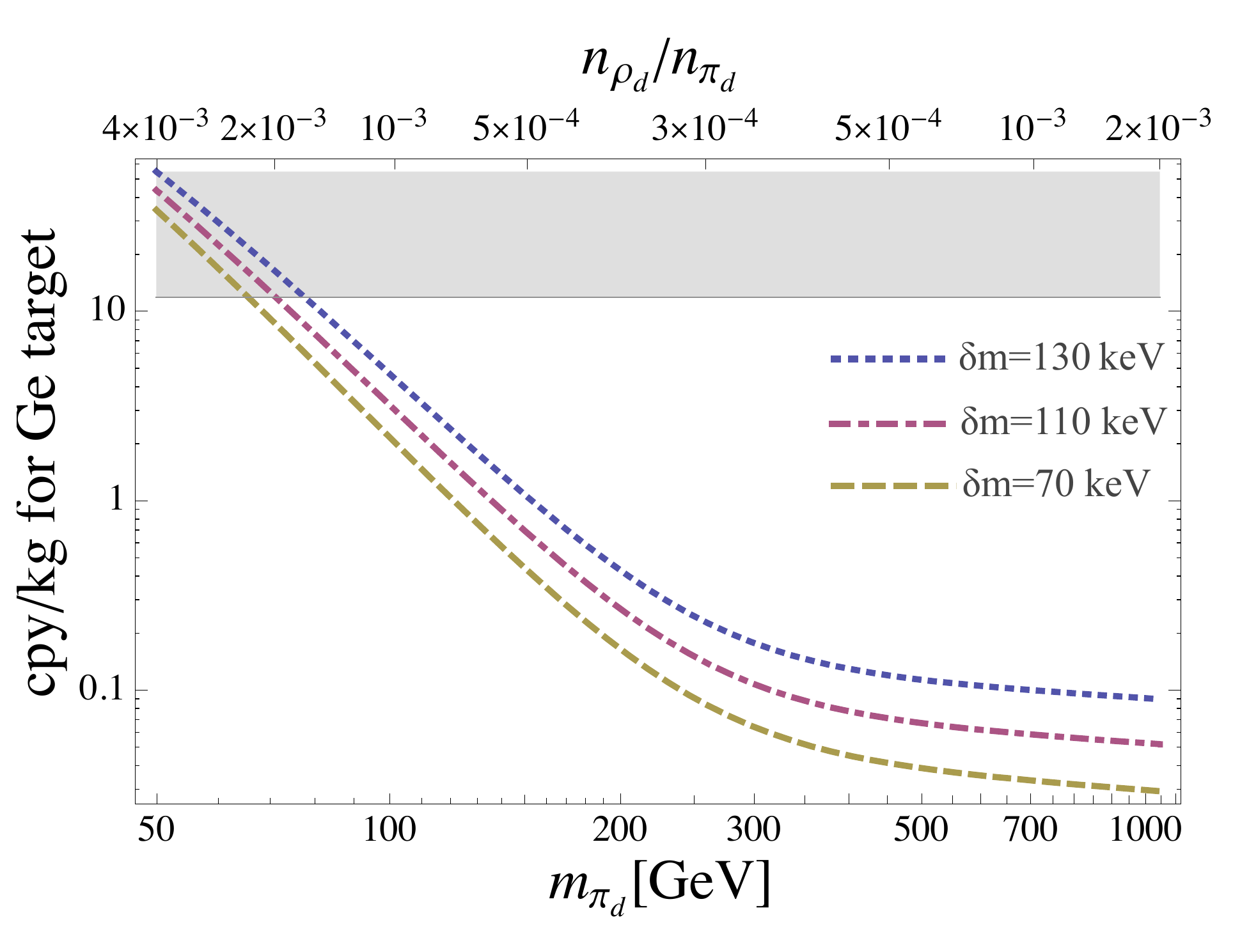}
\caption{Predicted number of events per kg-yr of exposure due to $\rhod\rightarrow\pid$ down-scattering off of a Ge target, and detector sensitivity in the 10-100 keV range. The gray region is excluded by the CDMS 2009 Ge analyzed data \cite{Ahmed:2009zw} at \mbox{95\% C.L.}}
\label{Fig:SpinTemp}
\end{center}
\end{figure}

The differential rate per unit detector mass is given by:
\begin{equation}
\label{Eq:diff cross-section}
\frac{dR}{dE_R}=\frac{n_\rhod}{n_\pid} \frac{\rho^0_{\pid}}{m_\pid m_N}\int_{v_{\text{min}}}\hspace{-0.15in}d^3\vec{v}\;\; v f(\vec{v},\vec{v}_E)\frac{d\sigma}{dE_R},
\end{equation}
where $\rho^0_\pid$ is on the order of the local dark matter density, $\rho^0_{\text{DM}}\approx 0.3\text{ GeV/cm$^3$}$, since we are assuming here that $\pid$ is the dominant halo component.
The Standard Halo Model (SHM)  velocity distribution function $ f(\vec{v},\vec{v}_E)$ with a functional form
\begin{equation}
 f(\vec{v},\vec{v}_E)=\mathcal{N}\left(e^{-(\vec{v}+\vec{v}_E)^2/v_0^2}-e^{-v^2_{\text{esc}}/v^2_0} \right)\Theta(v_{\text{esc}}-|\vec{v}+\vec{v}_E|)
\end{equation}
will be used in this article, with the following benchmark halo parameters: local escape velocity $v_{\text{esc}}=550\text{km/s}$, local velocity dispersion $v_0=270\text{km/s}$, and the velocity of the Earth in the galactic frame $\vec{v}_E$ as in \cite{Cui:2009xq}. Moreover, the minimum relative velocity for $\rhod$ to inelastically down-scatter to $\pid$ causing the target nucleus to recoil with energy $E_R$ is given by
\begin{equation}
v_{\text{min}}(E_R)=\frac{1}{\sqrt{2m_N E_R}}\left|-\delta m+\frac{E_Rm_N}{\mu}\right|,
\end{equation}
where $\mu$ is the DM-nucleus reduced mass.

%This simple kinematical fact has two interesting consequences: (i) the rate is suppressed at low energies, since the hyperfine splitting energy is converted to kinetic energy after de-excitation, and
%(ii) the annual modulation phase is inverted relative to the $\pid\rightarrow\rhod$ rate, \textit{i.e.}, the rate peaks on \mbox{Dec. $2^\text{nd}$.} Nevertheless, the modulation fraction is of order of a
%few percent, as for elastic scattering, since most of the halo is kinematically allowed to scatter.

Figure \ref{Fig:SpinTemp} displays the predicted rate at a detector with Ge as the nuclear target, with sensitivity in the range $10-100\keV$, such as CDMS. The region favored by DAMA/LIBRA evades all current bounds, but the next generation of direct detection experiments, such as XENON100, should have sensitivity to discover it. The distinguishing features of such down-scattering signals that can help disentangle it from other components are:  (i) lower recoil energy threshold below which the signal vanishes, since the hyperfine splitting energy is converted to kinetic energy after de-excitation, and (ii) for the region where the signal peaks, around $E_R=\delta m \frac{m_\pid}{m_N}$, there is a phase inversion in the annual modulation relative to the up-scattering rate (\textit{i.e.}, the $\rhod\rightarrow\pid$ rate peaks on \mbox{Dec. $2^\text{nd}$}) \cite{Graham:2010ca}. Nevertheless, the modulation fraction is of order of a few percent, as for elastic scattering, since most of the halo is kinematically allowed to scatter.

%%%%%%%%%%%%%%%%%%%%%%%%%%%%%%%%%%%%%%%%
\section{Baryon Elastic Component and Direct Detection}
\label{Sec:Baryons}

The calculations of Sec.~\ref{Sec: DMSynth} indicate that it is possible to have significant synthesis of the $\pid$ into baryons.
When this occurs, there are equal numbers of heavy baryons and light antibaryons.  This section considers the direct detection implications of baryons that can arise through synthesis in the early Universe. 
These baryons have large spin since they consist of single flavor quarks, and the axial constituent coupling will lead to a small but potentially detectable elastic scattering channel.

\subsection{Heavy Baryon Scattering}
\label{Sec: Heavy Baryon} 

As discussed in Sec.~\ref{FSSpec}, a $B_H$ heavy baryon has no hyperfine splitting since all its constituents have the same flavor. Hence its dominant scattering channel is elastic. If its constituents were vectorially coupled to the dark group $U(1)_\td$, dark heavy baryons would have an upper limit on their density fraction set by null searches (such as CDMS or XENON10) to be in the range $10^{-4}-10^{-8}$ (assuming that the inelastic component fits the DAMA signal). That would imply that CiDM would have to live in Region V of the $\Lambda_\td - m_{\text{light}}$ parameter space as described in Sec. \ref{Sec: PostSynth}; all other regions being excluded.

The picture changes when the dark quarks have a purely axial coupling to $U(1)_\td$. 
We shall demonstrate that in that case the $U(1)_\td$ gauge boson couples to the spin of the heavy baryon in the non-relativistic limit. 
This leads to a suppressed scattering rate relative to the case where the baryon has a vector-like $U(1)_\td$ charge, opening up the parameter space to allow for all processing regions described in Sec. \ref{Sec: PostSynth}. One remarkable consequence of this framework is the possibility that the dominant component of the halo is in the form of dark heavy baryons and what the DAMA/LIBRA experiment is detecting is a subdominant inelastic component of the halo. In this section we will make these statements quantitative and consider their consequences for direct detection.

The fermionic axial current is given by
\begin{equation}
J^\mu_{\text{A}}=\sum_{n}q^n_{\text{A}} \overline{\Psi}_n\gamma^\mu\gamma^5\Psi_n,
\end{equation}
where the index $n$ sums over constituent fermions and $q^n_{\text{A}}$ refers to their axial charge. Taking the non-relativistic limit and integrating out the small components,  the expression above for $N_c$ heavy quarks reduces to:
\begin{eqnarray}
\label{Eq:axial current}
J^\mu_{B_H}
&=&q_{\text{A}} \delta^\mu_i \chi^{\dagger}_{B_H} \mathbf{S}^i_{B_H}\chi_{B_H}+\mathcal{O}(1/m_H),
\end{eqnarray}
where $\chi_B$ is the $N_c/2$-spin wave function of the heavy baryon bound state,
\begin{eqnarray}
\chi_{B_H}=\psi_1\otimes\psi_2\otimes...\otimes\psi_{N_c}
\end{eqnarray}
\textit{i.e.} the direct product of the two component constituent wave functions $\psi_n$. The heavy baryon spin operator $\mathbf{S}^i_{B_H}$ is given in terms of the constituent spin operators as
\begin{eqnarray}
\mathbf{S}^i_{B_H}= \sigma^i_1\otimes\mathbf{1}_2\otimes...\otimes\mathbf{1}_{N_c}+
\ldots+\mathbf{1}_1\otimes...\otimes\mathbf{1}_{N_c-1}\otimes\sigma^i_{N_c}.
\end{eqnarray}
Note that its Casimir invariant is given by
\begin{equation}
\mathbf{S}_{B_H}^2=\frac{N_c}{2}\left(\frac{N_c}{2}+1\right),
\end{equation}
and it satisfies the following algebraic relations
\begin{eqnarray}
[\mathbf{S}^i_{B_H},\mathbf{S}^j_{B_H}]&=i\epsilon^{ijk}\mathbf{S}^k_{B_H}
\qquad
\{\mathbf{S}^i_{B_H},\mathbf{S}^j_{B_H}\}&=\frac{1}{6}N_c(N_c+2)\delta^{ij}.
\end{eqnarray}

The differential cross section for a dark heavy baryon scattering off a target nucleus
can be computed using Eq.~\ref{Eq:axial current} and the kinetic mixing between  $U(1)_\td$ and $U(1)_Y$, recalling that this results in the proton picking up an effective $\epsilon e$ charge under $U(1)_\td$.  The resulting cross section is
\begin{equation}
\frac{d\sigma}{dE_R}=\frac{N_c(N_c+2)}{12(N_c+1)}\frac{\alpha Z^2}{f^4_{\text{eff}}}\frac{1}{v_{\text{rel}}^2}E_R|F(E_R)|^2,
\end{equation}
and the differential rate per unit detector mass is given by:
\begin{equation}
\label{Eq:diff cross-section}
\frac{dR}{dE_R}=F_{\rho_{B}} \frac{\rho_{\text{DM}}^0}{m_{B_H}m_N}\int_{v_{\text{min}}}\hspace{-0.15in}d^3\vec{v}\;\; v f(\vec{v},\vec{v}_E)\frac{d\sigma}{dE_R}.
\end{equation}
Here, $F_{\rho_{B}}\equiv\rho_{B_H}/\rho^0_{\text{DM}}$ is the fractional dark matter density in the form of heavy baryons and $m_{B_H}=N_c m_H$ is the dark heavy baryon mass. The velocity distribution function we use is the same as in the previous section.   $v_{\text{min}}$ denotes the minimum relative velocity for a dark heavy baryon to elastically scatter off a nucleus with recoil energy $E_R$
\begin{equation}
v_{\text{min}}(E_R)=\frac{\sqrt{2m_N E_R}}{2\mu},
\end{equation}
where $\mu$ is the $m_{B_H}$-nucleus reduced mass.

\begin{figure}[htbp]
\begin{center}
\includegraphics[width=4in]{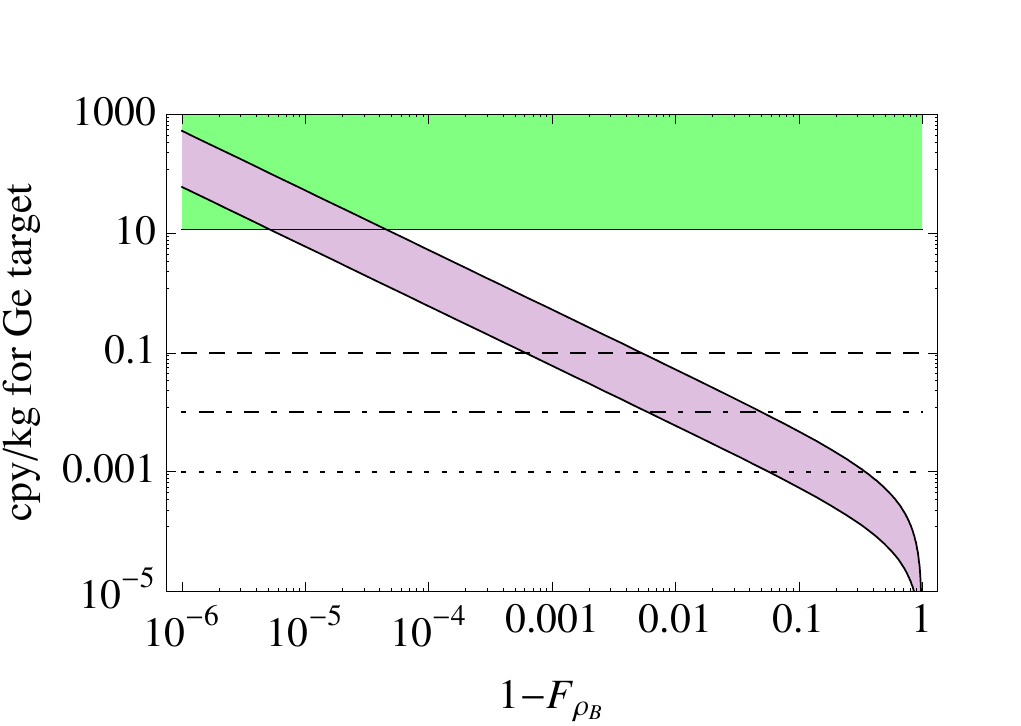}
\caption{Expected number of events per kg-yr from elastic heavy baryon scattering off Ge for $m_H=72\GeV$ and nuclear recoil range \mbox{10-100 keVnr}, taking the inelastic halo component to fit the DAMA signal. $F_{\rho_B}$ denotes the fractional dark matter density in the form of heavy baryons. The green region is excluded by the CDMS 2009 Ge analyzed data \cite{Ahmed:2009zw} at \mbox{95\% C.L}. The dashed line denotes the projected sensitivity of SuperCDMS for \mbox{30 kg-yr} of exposure at Soudan mine, the dot-dashed line \mbox{300 kg-yr} at Snolab and the dotted line \mbox{3 ton-yr} at DUSEL/GEODM.}
\label{Fig:CDMScounts}
\end{center}
\end{figure}

Fig.~\ref{Fig:CDMScounts} illustrates how the number of events expected at a Ge detector with sensitivity in the $10-100 \keV\text{nr}$ nuclear recoil range (such as CDMS) depends on the fractional heavy baryon density for $m_H=72\GeV$, assuming that the inelastic component fits the DAMA/LIBRA signal \cite{Lisanti:2009vy}. Note that if the halo is highly dominated by heavy baryons, with only a fraction of $10^{-3} - 10^{-5}$ in other components, then detection of the heavy baryon component is around the corner  \cite{Ahmed:2009zw}.

\subsection{Light Baryon Scattering}
\label{Sec: Light Baryons}

Recently, CoGeNT has seen anomalous low energy events consistent with light dark matter ($m\sim 7 \GeV$) elastically scattering \cite{Cogent, Cogent1, Cogent2, Cogent3}.  The rate is slightly larger than DAMA's rate if channelling is included\cite{Channelling, Chang:2008, Channelling0, Channelling1,Channelling2, Channelling3}.  A candidate for such light dark matter inside CiDM are the light baryons that arise while synthesizing the dark pions in heavy baryons.    There are two issues with this interpretation, the first is that CoGeNT's rate would oversaturate the light baryon rate at DAMA's detector obviating the need for inelastic dark matter. However, channeling at low nuclear recoil energies is uncertain and may not be as effective as naive extrapolations from higher nuclear recoil energies indicate\cite{Blocking,LowerChannelling1, LowerChannelling2, LowerChannelling3,LowerChannelling4}.  If channeling is less effective, then DAMA's signal may still arise from inelastic scattering, leaving CoGeNT's signal unexplained.

The second, more serious problem with explaining CoGeNT's anomaly with light baryons is that there are equal numbers of heavy baryons produced.  These heavy baryons have approximately the same  per nucleon cross section as the light baryons.   Heavier mass dark matter candidates have tight limits on their elastic scattering.

There are two possibilities to evade the heavy baryon limits.  Heavy baryons can decay into a light baryons and other colored particles.  These types of decays generically result from the interactions that generate the  asymmetry responsible for the $\pid$ abundance.     It is possible that the heavy baryons decay while leaving the $\pid$ stable resulting in a dark matter sector dominated by dark mesons and light baryons.  For instance,  if the dominant operator that violates $H-L$ number has charge 4, \eg
\begin{eqnarray}
\LL_{H-L \text{violation}} = \frac{1}{M^2} H H L^c L^c,
\end{eqnarray}
then $\pid$ will be stable, all other heavy dark hadrons will be unstable.  Specifically, the baryons will chain decay as
\begin{eqnarray}
B_H \rightarrow B^{(N_c-2)} +X\rightarrow \cdots \rightarrow   \begin{cases} B_{\bar L} + X& N_c \text{ even}\\ \pidn{1} + B_{\bar L} +X & N_c \text{ odd}\end{cases}.
\end{eqnarray}
  While these decays  are model dependent, it might be possible to correlate cosmic ray signals from dark meson decay or oscillation with CoGeNT's signal.  
  
Another method for evading the heavy baryon constraints is to have a non-zero $H+L$ number generated resulting in an excess of $B_{\bar L}$'s.  
 Generating a significant dark baryon asymmetry for large $N_c$ is challenging because the dark baryon operators have high scaling dimension and therefore are highly irrelevant.  For $N_c =3, 4, 5$ it may be possible to generate a sizeable asymmetry, particularly in supersymmetric theories because dark squarks have lower scaling dimension than their fermionic partners.

%%%%%%%%%%%%%%%%%%%%%%%%%%%%%%%%%%%%%%%
\section{Conclusion}\label{Sec:Conclusion}

In Composite inelastic Dark Matter, the $\OO(100\keV)$ mass splitting suggested by DAMA/LIBRA arises from dynamics inside composite states of a strongly interacting sector. 
Dark hadrons carrying non-zero flavor quantum numbers are stable and hence potential dark matter candidates. In particular, heavy flavor mesons offer a good candidate for implementing iDM. 
The thermal relic abundance of dark hadrons is too small to account for all the dark matter due to strong self-interactions, meaning that the dark matter abundance in these theories must originate from a dark flavor (or dark baryon) asymmetry.
One of the purposes of this work was to investigate the cosmological evolution of the flavor asymmetry and how it determines the mesonic and baryonic abundances at late times. 
This question was addressed in the context of the minimal model of Sec.~\ref{intro:summary}, where the iDM candidate is a spin-0 dark meson $\pid$ that inelastically scatters to a nearly degenerate vector state $\rhod$.

In a large part of the parameter space, $\pid$ mesons dominate the abundance of dark matter. 
In other regions, dark baryons dominate the abundance; however, the residual $\pid$ component interacts sufficiently strongly to give rise to a viable iDM scenario. In all cases, exotic dark matter components arise with novel elastic scattering properties -- nuclear recoil events are suppressed at low-energy by a dark matter form factor. The relative abundance of $\rhod$ to $\pid$ is typically $\sim 10^{-4}$, suggesting that $\rhod \rightarrow \pid$ down scattering off nuclei is a discoverable signal in the near future. 
Finally, we studied a variety of long lived meson and baryons states and found that BBN constraints on their decays are not severe.
 \vspace{0.2in}

%%%%%%%%%%%%%%%%%%%%%%%%%%%%%%%%%%%%%%%
\noindent
{ \bf Acknowledgements \vspace{0.05in}}

We would like to thank Mariangela Lisanti for helpful feedback as well as Rouven Essig and Natalia Toro for illuminating discussions.
DSMA, SRB, PCS  and JGW are supported by the US DOE under contract number DE-AC02-76SF00515.  JGW is supported by the DOE's Outstanding Junior Investigator Award.

\appendix

%%%%%%%%%%%%%%%%%%%

%%%%%%%%%%%%%%%%%%%%%%%
%%%%%%%%%%%%%%%%%%%%%%%%%%%%%

\end{document}